\documentstyle[12pt]{article}
\begin{document}

\baselineskip=8mm

\title{\bf Turbulence: from Complexity to Variety}
\author{M. Germano  \\ \\
\small Department of Civil and Environmental Engineering\\
\small Duke University, Durham, NC 27708, USA\\
\small mg234@duke.edu}

\date{}
\maketitle

\abstract{In this note we advocate the notion of variety as juxtaposed to the notion of complexity. Laminar flows are complex, turbulence is various. When the gradients reach a critical point, laminar flows are subjected to instabilities and transitions, sometime soft, sometime dramatic, and the final result is the variety. Turbulent flows in nature are fascinating, and loosely speaking we could say that a system is complex when differences are suffered, is various when diversities are enjoyed and exploited as opportunities.}

\section*{}

Scientific imagination feeds on similarities. Analogies between different problems, and metaphorical approaches are  powerful transmission belts between different disciplines. Nature offers us a fantastic laboratory for all that with the mechanics of fluids. Laminar flows are ordered and well defined, but when the gradients, or more correctly the Reynolds number, reach a critical value, instabilities occur and a new organization takes place. We call it turbulence, and strangely enough it is usual to refer to turbulence as a complex situation. But why Nature should be complex in the essence?

Definitions of complexity abound, but in many cases we simply define complex a situation difficult to understand and to reduce to our present knowledge. From the essential point of view we personally think that the notion of complexity should be conjugated jointly with the notion of variety. A system is complex when differences are suffered, is various when diversities are enjoyed and exploited as opportunities. When the gradients reach a critical point, complex systems are subjected to instabilities and transitions, sometime soft, sometime dramatic, and the final result is the variety.

Turbulence is not complex, is various. The ways in which turbulence puts to work eddies, vortices, coherent structures in order to realize a mysterious project is fascinating. The ability of its active variety to overcome obstacles, to mix species has something of divine;  we have a lot to learn from that. Turbulent flows are very important from the practical point of view and from the beginning a test bench for new mathematical representations and decompositions. Turbulence in fluids is characterized by a chaotic motion seeded with coherent structures, difficult to extract and to filter out. Fundamental is the problem of the average, both in its traditional statistical form and in the filtered one. Strangely enough the scientific interest for the complexity largely overcomes the interest for the variety. Definitions of complexity abound, definitions of variety are lacking. 

Why turbulence, the variety, is so difficult to understand? It is not so easy to say. In a book \cite{Iri77} Luce Irigaray remarks a historical negligence of science for the mechanics of fluids. The reason for her is that the masculine attitude has privileged the mechanics of solids, but Alan Sokal and Jean Bricmont \cite{Sok98} remark that we know the equations that describe the fluid motions: the problem is that they are difficult to solve, particularly as regards turbulent flows. Everything should be explained by the Navier-Stokes equations, but they are elusive and unresolved as regards the fundamentals. Existence and smoothness of the solutions remain unsolved millennium problems.

It is an old story, but let us discard for a moment the turbulence, let us consider another old story, the story of subsonic lift, the force that raises us and does not work, free of charge, we could say. If we open the book {\it A history of aerodynamics} \cite{And98} by John Anderson, we find at page three the following words: {\it Why were the principles of aerodynamics considered so mysterious, and why was it so difficult to develop them into an applied science?}   In an older paper \cite{Gia34} on the history of aerodynamics, we read that till to the beginning of the past century {\it of the two aspects in which air reaction appears, known today as lift and drag ... we have only to note researches on the second one}. We recall that the resistance of a plate normal to the flow at high Reynolds numbers was satisfactorily solved at the time of Newton, so why as regards the lift we have to remark this negligence? It is really so difficult for a teacher today to introduce, at least qualitatively, the physical reasons of subsonic lift?

No doubt that as regards the flight we have two parallel stories. The first, well known, is the story of people desperately in love with air, decided to fly against any {\it scientific} evidence, and convinced that the air could provide that. The Wright Brothers, Lilienthal, Ferber and many others had to excite personally their own physical imagination in order to produce the first flyers. The other story regards how we have rationally and quantitatively provided the mathematical tools for that, and here we have to admit a failure, a historical negligence, and in our opinion some reason for that negligence is concealed in the observations of Luce Irigaray. The air that resists, the air that must be penetrated, was offering free of charge on the wing an enormous force, if compared to drag, (ten to one, and more, in our modern airplanes), and nobody for a long time realized that {\it scientifically}. Hovering birds, sails, flags, kites were at our disposal, but till to the clear evidence of the first human flight nothing happened in our drag-conditioned scientific imagination.

In order to arrive to the Joukowsky theorem and to the formulation of the Kutta condition a new scientific sensibility had to emerge, and probably today we need something like that in the study of turbulence. Turbulence, and variety, require something new. Probably also in this case we have to remove prejudices, old ways of thinking, obsolete approaches. There is a lot of work to do.

\end{document}